\documentclass[twocolumn,prl]{revtex4}

\usepackage{graphicx}
\usepackage{dcolumn}
\usepackage{bm}
\usepackage{graphics}
\usepackage{mathrsfs}

\newcommand{\be}{\begin{equation}}
\newcommand{\ee}{\end{equation}}
\newcommand{\bea}{\begin{eqnarray}}
\newcommand{\eea}{\end{eqnarray}}

\begin{document}

\title{Electrostatic theory of viral self-assembly: a toy model}

\author{Tao Hu}
\author{Rui Zhang}
\author{B. I. Shklovskii}
\affiliation{Theoretical Physics Institute, University of
Minnesota, Minneapolis, Minnesota 55455}

\begin{abstract}
Viruses self-assemble from identical capsid proteins and their
genome consisting, for example, of a long single stranded (ss) RNA.
For a big class of T = 3 viruses capsid proteins have long positive
N-terminal tails. We explore the role played by the Coulomb
interaction between the brush of positive N-terminal tails rooted at
the inner surface of the capsid and the negative ss RNA molecule. We
show that viruses are most stable when the total contour length of
ss RNA is close to the total length of the tails. For such a
structure the absolute value of the total RNA charge is
approximately twice larger than the charge of the capsid. This
conclusion agrees with structural data.
\end{abstract}

\maketitle

Unlike living cells, viruses do not have any metabolic activity,
which may mean that they are in the state of thermal equilibrium.
This is one of the reasons why the statistical physics can be used
for understanding of viruses. The structure of viruses is also
dramatically simple. Inside the protein capsid each virus carries
its genome, which consists of one or more DNA or RNA molecules and
is used for reproduction in host cells. The focus of this letter is
on viruses with single stranded RNA (ss RNA) genomes. Detailed image
reconstruction of apparently spherical viruses reveals their
icosahedral symmetry. This is why such a virus capsid can be viewed
as a curved two-dimensional crystal closed on
itself~\cite{Enc,Annette,Bruinsma}.

\begin{figure}[htb]
\centering
\includegraphics[width=0.45 \textwidth]{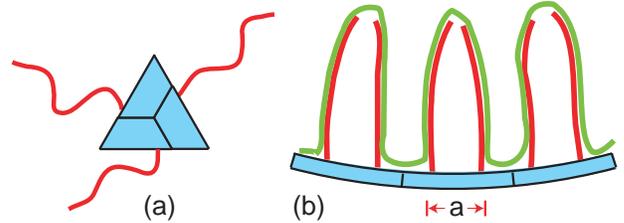}
\caption{(color online) Schematic sketch of the protein capsid
assembly. (a) Triangular block made of three proteins (blue) with
their positive flexible N-terminal tails (red). (b) The brush of
positive N-terminal tails rooted at the inner surface of the capsid
made of triangular blocks. The ss RNA (green) strongly interacts
with the tails and keeps all the blocks together. }
\label{fig:virus}
\end{figure}

Here we concentrate on the viruses of the so called T = 3 class,
in which a capsid is made of precisely 180 identical proteins, or
of 60 triangular blocks consisting of three proteins each (see
Fig. \ref{fig:virus}). In-vitro studies of solutions of capsid
proteins and RNA molecules of a given virus show that under the
biological pH and salinity they can spontaneously self-assemble
into infectious viruses~\cite{hung,Brancroft,adolph}. This letter
focuses on the energetics of this amazing protein-RNA
self-assembly. In addition to hydrophobic attraction between the
proteins it is driven by strong Coulomb attraction between capsid
proteins and RNA molecules~\cite{Brancroft,Bruinsma}. Indeed, ss
RNA is strongly negatively charged. Its backbone has one negative
phosphate per nucleotide or per $0.65$ nm. We denote the total ss
RNA charge of  a virus particle as $-Q_r$. According to Tab.
\ref{tab:charge} of T = 3 viruses $Q_r$ is about several thousand
in units of the proton charge. On the other hand, for many viruses
their capsid proteins carry substantial net positive charge $q_p$,
which can reach 17. The net positive charge of the capsid of a T =
3 virus $Q_{c} = 180 q_{p}$ can, therefore, reach 3000. Although,
in biological conditions the protein-RNA interaction is screened
by monovalent salt at the Debye-Huckel screening radius $r_s$,
attraction energy of such big charges is still very large.

A dramatic feature of the group A of T = 3 viruses collected in
the upper part of Tab. \ref{tab:charge} is that almost all the
capsid protein charge is concentrated in the N-terminal tail
located inside the capsid (Fig. \ref{fig:virus}). We define such
an N-terminal tail as the flexible sequence of amino acids, which
starts from the N-terminus of the protein and ends at the first
$\alpha$-helix or $\beta$-sheet. It looks like evolution created
cationic N-terminal tails for the strong interaction with ss RNA
genome (Fig. \ref{fig:virus}b).

In this letter we concentrate on the electrostatic interaction of
the ss RNA with the brush of tails of a group A virus (see Fig.
\ref{fig:virus}b). In particular we want to understand a
remarkable fact that for these viruses the absolute value of the
ss RNA charge $Q_r$ is substantially larger than the total charge
of the capsid $Q_c = 180 q_p$. The charge inversion ratios $R =
Q_r/Q_c$ for them are given in Tab. \ref{tab:charge}. They are
scattered with the median value 1.8. This raises a challenging
question whether such ratio can be obtained by minimizing free
energy of the virus~\cite{Bruinsma1} with respect to RNA length.
The positive answer to this question was recently given in the
framework of the simplest model where positive protein charges are
uniformly smeared on the internal surface of the capsid, while the
ss RNA is adsorbed on this surface as a negative
polyelectrolyte~\cite{Bruinsma1}. As we see from Tab.
\ref{tab:charge} capsid charges of all the group A viruses are
concentrated in the tails. That is why we suggest an alternative
model of virus self-assembly, namely adsorption of ss RNA on a
brush of flexible positive tails, rooted on a neutral surface.
Minimizing the free energy of such self-assembly with respect to
the total ss RNA length we arrive at the theoretical charge
inversion ratios $\Re$, which are quite close to the the factual
ones $R$.
\begin{table}
\caption{The absolute value of ss RNA charge $Q_r$, the charge of
the capsid protein $q_p$, the N-terminal tail charge $q_t$, the
number of amino acids in the tail $N_t$, the ratio of the linear
charge densities (in fully stretched state) of the ss RNA $\eta_r$
and the tail $\eta_t$, the ratio $N_d/N_t$, where $N_d$ is number of
amino acids in disordered part of the tail, the actual and predicted
charge inversion ratios $R$ and $\Re$. The data are obtained from
Refs.~\cite{Bank,Explorer}. In the group A most of the capsid
charges are concentrated in the tails. In the group B the protein
charges are large but the tails are practically neutral. In the
group C the charges of both capsid proteins and tails are very
small.} \label{tab:charge}
\begin{tabular}{|l|l|l|l|l|l|l|l|l|}
\hline Virus & $Q_r$ & $q_p$ & $q_t$ & $N_t$ & $\frac{\eta_r}{\eta_t}$ & $\frac{N_{d}}{N_{t}}$ & $R$ & $\Re$\\
\hline Group A\\
\hline Brome Mosaic & 3030 & 10 & 9 & 48 & 2.8 & 0.44 & 1.7 & 2.8\\
\hline Cowpea Chlorotic Mottle & 2980 & 7 & 9 & 49 & 2.8 & 0.63 & 2.4 & 2.8\\
\hline Cucumber Mosaic & 3214 & 15 & 12 & 66 & 2.9 & 0.59 & 1.2 & 2.9\\
\hline Tomato Aspermy & 3391 & 14 & 12 & 67 & 2.9 & 0.55 & 1.3 & 2.9\\
\hline Nodamura & 4540 & 13 & 18 & 52 & 1.5 & 0.79 & 1.9 & 1.8\\
\hline Pariacoto & 4322 & 13 & 14 & 47 & 1.8 & 0.61 & 1.8 & 1.9\\
\hline Sesbania Mosaic & 4149 & 6 & 6 & 57 & 5.0 & 0.89 & 3.8 & 5.0\\
\hline Rice Yellow Mottle & 4450 & 17 & 12 & 52 & 2.3 & 0.79 & 1.5 & 2.3\\
\hline Southern Bean Mosaic & 4136 & 16 & 14 & 58 & 2.2 & 0.89 & 1.4 & 2.2\\
\hline Cocksfoot Mottle & 4082 & 15 & 13 & 54 & 2.2 & 0.88 & 1.5 & 2.2\\
\hline Carnation Mottle & 4003 & 10 & 11 & 81 & 3.9 & 1.0 & 2.2 & 3.9\\
\hline Tobacco Necrosis & 3700 & 9 & 10 & 79 & 4.1 & 0.90 & 2.3 & 4.1\\
\hline Tomato Bushy Stunt & 4776 & 12 & 13 & 92 & 3.7 & 0.91 & 2.2 & 3.7\\
\hline Group B\\
\hline Dengue & 10735 & 9 & 1 & 7\\
\hline Immature Yellow Fever & 10862 & 14 & 1 & 7\\
\hline Immature Dengue-2 prM & 10703 & 15 & 1 & 7\\
\hline Group C\\
\hline Norwalk & 7654 & 4 & -2 & 28\\
\hline Native Calicivirus & 7500 & 1 & -4 & 26\\
\hline Bacteriophage Q Beta & 4215 & 3 & 1 & 3\\
\hline Bacteriophage Ga & 3466 & 3 & 0 & 3\\
\hline Bacteriophage MS2 & 3569 & 1 & 0 & 4\\
\hline Bacteriophage Fr & 3575 & 1 & 0 & 4\\
\hline Bacteriophage Pp7 & 3588 & 1 & 0 & 3\\
\hline Bacteriophage alpha3 & 6087 & -4 & -1 & 8\\
\hline Turnip Yellow Mosaic & 6318 & 0 & -2 & 29\\
\hline Desmodium yellow mottle & 6300 & 5 & 1 & 29\\
\hline Physalis Mottle & 6673 & 4 & -1 & 28\\
\hline
\end{tabular}
\end{table}
\begin{figure}[htb]
\centering
\includegraphics[width=0.45 \textwidth]{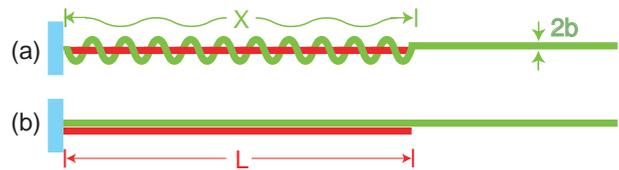}
\caption{(color online) Complexes of the long ss RNA (green) with
a cationic tail (red) rooted on the internal surface of capsid
(blue). $L$ is the length of the tail, $X$ is the length of RNA
piece, which complexes with the tail. The structure and the
magnitude of $X$ depends on the ratio between the charge densities
of the tail and the ss RNA. (a) $X > L$, when $\eta_r< 2\eta_t$;
(b) $X = L$, when $\eta_r>2\eta_t$.} \label{fig:theory}
\end{figure}

We call our model a toy model because we start from the following
two simplifications. (i) First, similar to Ref.~\cite{Bruinsma1} we
neglect hydrogen bonds between ss RNA bases which lead to the
secondary structure of ss RNA. (ii) Second, we assume that each tail
is free (does not stick to the capsid surface). Actually for some
tails, their part close to the tail root sticks to the capsid
surface~\cite{book}. Only this part of the N-terminal tail is seen
in the X-ray images of the crystallized viruses, while the rest of
the tail is missing. Missing part of the tail strongly fluctuates
and is called disordered. We call $N_d$ the average number of amino
acids in the disordered (free) part of the tail. Ratios $N_d/N_t$
are given in Tab. \ref{tab:charge}. We see that on average $76\%$ of
the tail length is free. In our toy model we assume that $N_d =
N_t$.

Let us first consider interaction of a homo-polymeric ss RNA with
a single free cationic N-terminal tail rooted at the neutral
internal surface of the capsid (Fig. \ref{fig:theory}a). We assume
that in fully stretched state each tail has length $L$ and the
positive linear charge density $\eta_t$, while the very long ss
RNA  in fully stretched state has the negative linear charge
density $-\eta_r$. The ss RNA piece of the length $X \ge L$
complexes with the tail. Both polymers are modelled as worm-like
chains with the same radius $b$, which is simultaneously of the
order of their bare persistence length $p_0$ (which does not
include Coulomb self-repulsion). The third important assumption of
our toy model is that (iii) the solution has a moderate salt
concentration, so that $b \ll r_s \ll L$. We argue below that even
this assumption does not change our results qualitatively.

Due to the strong Coulomb repulsion inside the overcharged
complex, the strongly negatively charged ss RNA has a relatively
large persistence length $p \sim r_{s}^{2}/p_0$ (see
Refs.~\cite{Od,Fix,Toan}), so that its Coulomb energy can be
estimated as the energy of a rigid cylinder of the radius $b$.
Same is true for the complex of the N-terminal tail and the ss
RNA, which as we will see has the large negative linear charge
density $\eta^{*}$. Self-repulsion of these negative charges makes
the complex locally stretched, so that its total length equals
$L$. Therefore, $\eta^{*} = (-X\eta_r + L\eta_t)/L$. The tail-RNA
complex with the long ss RNA shown in the Fig. \ref{fig:theory}a
has the large electrostatic energy. Therefore, the contribution to
the free energy $F$ from configurational entropy plays a minor
role and can be neglected. Since $r_s \ll L \le X$, the Coulomb
interaction is truncated at $r_s$. As a result, we obtain the
following simple expression for the $X$-dependent part of the free
energy
\be F(X) =
L\left(\frac{-X\eta_r+L\eta_t}{L}\right)^2\ln\left(\frac{r_s}{b}\right)-X\eta_r^2\ln\left(\frac{r_s}{b}\right).
\label{eq:energy} \ee
The first term represents the self-energy of the overcharged
N-terminal tail (the complex), while the second term represents
the loss of the electrostatic energy of the ss RNA segment with
length $X$. Here we neglect the Coulomb repulsion between the
complex and the rest of the ss RNA because $r_s \ll L$, $X$.
Minimizing $F(X)$ with respect to $X$, we find the optimal $X =
X_0 = (\eta_t/\eta_r+1/2)L$, and the linear charge density of the
complex $\eta^{*} = -\eta_r/2$~\cite{foot00}. As we expected,
$\eta^{*}$ is negative, so the N-terminal tail is overcharged by
the ss RNA. The above calculation is valid if ss RNA wraps around
the tail (Fig. \ref{fig:theory}a) and, therefore, $X_0 > L$. This
happens only at $\eta_r/\eta_t < 2$. On the other hand at
$\eta_r/\eta_t = 2$, the length of the ss RNA segment in the
complex, $X_0$ reaches the minimum possible value $X_0 = L$
corresponding to stretched ss RNA. At $\eta_r/\eta_t > 2$ both
polymers are stretched (Fig. \ref{fig:theory}b) by the Coulomb
self-repulsion, $X_0 = L$, and $\eta^{*}= \eta_t - \eta_r <
-\eta_t$. Thus, at $\eta_r/\eta_t > 2$ the tail is overcharged by
ss RNA more than twice.

Until now we assumed that the ss RNA length $\mathscr{L}$ is always
larger than $X_0$, so that $X_0$ does not depend on $\mathscr{L}$.
Let us now imagine that we vary $\mathscr{L}$ at fixed $L$, $\eta_t$
and $\eta_r$. Then for a short ss RNA, $\mathscr{L} < X_0$, (where
$X_0$ is still the optimum value of $X$ found above) the new optimum
value of $X=X_{00}$ equals $\mathscr{L}$ (the N-terminal tail
consumes all available ss RNA). This means that at $\mathscr{L} <
X_0$ the electrostatic energy decreases with growing $\mathscr{L}$,
while for $\mathscr{L} > X_0$ the energy saturates. Thus, complex of
ss RNA with an N-terminal tail is most stable if $\mathscr{L} \geq
X_0$.

Now we can switch from a single N-terminal tail to the whole brush
of 180 tails and a very long ss RNA with the length $\mathscr{L}$
comparable to $180L$. The average distance $a$ between two
neighboring tail roots (see Fig. \ref{fig:virus}b) is typically
close to $5$~nm. We deal with $r_s$ much smaller than $a$, so that
complexes of the nearest neighbor  tails with RNA can be treated
separately. This means that long enough ss RNA goes from one tail
to another consequently overcharging each of them in the way we
calculated above for a single tail (Fig. \ref{fig:virus}b).

It is easy to show that if $\mathscr{L} < 180 X_0$ ss RNA is
shared between tails in equal portions $\mathscr{L}/180 < X_0$. In
this case the total electrostatic energy still goes down with
growing $\mathscr{L}$. (Here and below we neglect the length of ss
RNA per tail necessary to connect the tail roots: it is of the
order of $a/2 \ll L$. Indeed, according to Tab.~\ref{tab:charge}
$L \sim 15$ nm, while $a/2 \sim 2.5$ nm.) On the other hand, when
$\mathscr{L} > 180X_0$ and each N-terminal tail gets the length
$X_0$ of ss RNA, the electrostatic energy saturates at low level
and does not depend on $\mathscr{L}$. At this point in order to
find optimal length of ss RNA for given tails, we should recall
the excluded volume interaction energy, which is smaller than the
electrostatic energy, but provides the growth of the free energy
with $\mathscr{L}$ at $\mathscr{L} > 180X_0$. Indeed, one should
take into account that due to screening the persistence length of
the tail-RNA complex is much smaller than the tail length $L$ and
the tail-RNA "arches" are not extended as shown in Fig.
\ref{fig:virus}b, but rather tend to make coils. This leads to a
noticeable excluded volume interaction. Thus, for given tails the
free energy reaches minimum at $\mathscr{L} \simeq 180X_0$.
(Similar minimum was obtained earlier for the model of protein
charges uniformly smeared on the internal capsid
surface~\cite{Bruinsma1}.) For the theoretical charge inversion
ratio $\Re$ we arrive at
\be \Re = \frac{X_0\eta_r}{L\eta_t} = \left\{\begin{array}{lcr} 1+\eta_r/(2\eta_t), & {\rm when} & \eta_r < 2\eta_t \\
\eta_r/\eta_t, & {\rm when} & \eta_r > 2\eta_t
\end{array} \right. \ . \label{eq:R} \ee
In Tab. \ref{tab:charge} we calculated the ratio $\eta_r/\eta_t$
for the group A viruses using 0.65 nm for the distance between two
charges of ss RNA and 0.34 nm for a length of the tail per amino
acid. We see that for the most of the viruses $\eta_r/\eta_t \geq
2$ and, therefore, ss RNA is stretched along the N-terminal tails
(Fig. \ref{fig:virus}b), so that a simple way to formulate our
results for the length of ss RNA is to say that the total length
of ss RNA $\mathscr{L}$ is equal to the total length of the tails
$180L$. Substituting values of $\eta_r/\eta_t$ from Tab.
\ref{tab:charge} in Eq. (\ref{eq:R}) we arrived at values of $\Re$
listed in Tab. \ref{tab:charge}. We see that most of them are in
reasonable agreement with the structural data~\cite{foot0}.

This agreement may be interpreted as a result of natural evolution
of viruses in the direction of the maximum viral stability. It is
desirable, however, to design an in vitro experiment, which
verifies our predictions. Before suggesting such experiment let us
note that although above we discussed only packaging of a single
ss RNA molecule in a virus, our conclusions can be extended to the
case, where many shorter ss RNA pieces are packaged in the virus.
They just continue each other inside the virus and bind proteins
together. Our predictions, therefore, can be verified by
experiments with a solution of relatively short homo-polymeric ss
RNA with the length $\mathscr{L}$ in the range $2L < \mathscr{L}
\ll 180L$. We suggest an equilibrium experiment with a series of
solutions, which have a varying ratio $\rho$ of the total charges
of short ss RNA and capsid capsid proteins. At $\rho \simeq 1$ in
equilibrium all ss RNA molecules are used up in viruses, so that
there is no free ss RNA. With growing $\rho$ free ss RNA should
appear at the critical point $\rho = \rho_c = \Re$, where,
according to our theory, free ss RNA molecules and ss RNA
molecules inside the virus are in equilibrium. Using short ss RNA
permits to vary amount of ss RNA in a virus almost continuously in
order to find $\rho_c$ and compare it with $\Re$.

Let us now discuss the assumptions (i), (ii) and (iii) of our toy
model, starting from the assumption (ii), that one can treat the
N-terminal tail with a part of it sticking to the internal capsid
surface as a free tail. The picture of RNA going along the one side
of the tail without wrapping does not seem to be too sensitive to
the fact that the other side of the tail sticks to the capsid. This,
(together with the fact that in average only $24\%$ of the tail
length sticks to the capsid surface) makes (ii) reasonable. The
assumption (iii) is more problematic because biological values of
$r_s \sim b \sim 1~$nm. They easily satisfy inequalities $r_s \ll L,
a$, but do not literally satisfy assumption that $r_s \gg b$. This
assumption was important in order to say that ss RNA and N-terminal
tail-RNA complex are stretched and the Coulomb energy dominates the
configuration entropy. We argue here that according to numerical
simulations~\cite{Toan} for a very flexible polyelectrolyte (with
the bare persistence length equal to the Bjerrum length) even for
such a small $r_s$ the Coulomb interaction plays a strong role: its
persistence length grows three times already at $r_s = 1~$nm. For
less flexible polyelectrolyte such as ss RNA or the tail-RNA complex
the Coulomb interaction should play even stronger role so that for
zero order approximation the configuration of the complex shown in
Fig. \ref{fig:theory}b is reasonable. The assumption (i) that ss RNA
behaves as a flexible linear polyelectrolyte is not necessary for a
homo-polymeric ss RNA or a generic linear polyelectrolyte used for
virus self-assembly in-vitro~\cite{Brancroft}. On the other hand,
for the viral ss RNA, the energy of hydrogen bonds should be
optimized together with the electrostatic energy. It seems that
effect of such global optimization will not differ much from our
result, but this remains to be shown.

Up to now we have dealt with the group A. In the group B charges of
the capsid proteins are large but tails are practically neutral so
that the theory of Ref.~\cite{Bruinsma1} is appropriate. In the
group C the charges of proteins and tails are very small but it is
possible that for some viruses the internal surface of capsid
proteins is positively charged, while the negative charges are on
the external surface~\cite{footnote1}. In this case, one may also
redefine $R$ as ratio of ss RNA charge to the total charge of the
internal surface of the capsid and use Ref.~\cite{Bruinsma1} to
estimate $R$.

In this paper we focused on T=3 viruses, because they attract most
of physicists attention~\cite{Bruinsma,Bruinsma1,Bruinsma2}. As we
saw many of their capsid proteins have long positive tails. Capsid
proteins of some T=1, 4 and 7 viruses also have positively charged
tails. Our theory is applicable to them as well. Detailed analysis
of these classes is beyond scope of this paper.

In conclusion, the data~\cite{Bank,Explorer} show that there is a
big group of viruses, where practically all positive charges of a
capsid protein are concentrated in a long and flexible N-terminal
tail. For a given length and charge of the tail we optimized the
length of the ss RNA genome by searching for minimum of free energy
of the virus. We arrived at the very simple result that a virus is
most stable when the total length of ss RNA is close to the total
length of the tails. This result is in reasonable agreement with the
viral structural data~\cite{Bank,Explorer}. This may be interpreted
as a result of evolution in the direction of viral stability.

We are grateful to M. Rubinstein for valuable contribution during
initial stage of this work and to Z. Wang and M. Rubinstein for
showing us their unpublished computer simulations of similar to
Fig. 2a confirmations of diblock polymer molecule consisting of
two oppositely charged blocks. We appreciate useful discussions
with R. Bruinsma, A. Yu. Grosberg. T. T. Nguyen, and I. Rouzina.
B. I. S. acknowledges hospitality of Santa Barbara KITP and Aspen
Center of Physics, where this work was started.



\begin{thebibliography}{100}
\bibitem{foot00} For similar but not identical problem of flexible
polyelectrolyte wrapping oppositely charged rigid cylinder, this
result was previously obtained in Ref.~\cite{NgSh}.
\bibitem{foot0} One can see strong deviations from our predictions
for viruses in the third, fourth and the last three lines of the
group A. They have anomalously long tails with strongly non-uniform
distribution of charges in the tail~\cite{Bank,Explorer}. Heavily
charged domains occupy roughly speaking only half of the tail
length. Because of the salt screening neural pieces are not
stretched by the Coulomb repulsion of the neighboring charged
domains and acquire coil-like configurations. As a result, a charged
tail has twice larger linear charge density. Therefore, according to
Eq. (\ref{eq:R}) predicted $\Re$ should be roughly speaking twice
smaller, in a reasonable agreement with $R$. In essence in these
cases, we can not neglect the configurational entropy.
\end{thebibliography}
\end{document}